\documentclass{article} % For LaTeX2e
\usepackage{iclr2022_conference,times}

\usepackage{amsmath,amsfonts,bm}
\newcommand\Tstrut{\rule{0pt}{2.3ex}}
\newcommand\Bstrut{\rule[-0.9ex]{0pt}{0pt}}

\newcommand{\MPN}{\mathrm{MPN}}
\newcommand{\FFN}{\mathrm{FFN}}

\newcommand{\pad}{\mathtt{\langle MASK \rangle}}
\newcommand{\set}[1]{\{#1\}}

\def\eqref#1{equation~\ref{#1}}
\def\1{\bm{1}}

\def\vb{{\bm{b}}}
\def\vc{{\bm{c}}}
\def\vd{{\bm{d}}}
\def\ve{{\bm{e}}}

\def\vg{{\bm{g}}}
\def\vh{{\bm{h}}}

\def\vn{{\bm{n}}}

\def\vp{{\bm{p}}}
\def\vq{{\bm{q}}}

\def\vs{{\bm{s}}}

\def\vu{{\bm{u}}}
\def\vv{{\bm{v}}}

\def\vx{{\bm{x}}}

\def\mO{{\bm{O}}}

\def\mU{{\bm{U}}}
\def\mV{{\bm{V}}}
\def\mW{{\bm{W}}}

\DeclareMathAlphabet{\mathsfit}{\encodingdefault}{\sfdefault}{m}{sl}
\SetMathAlphabet{\mathsfit}{bold}{\encodingdefault}{\sfdefault}{bx}{n}

\def\gD{{\mathcal{D}}}
\def\gE{{\mathcal{E}}}

\def\gG{{\mathcal{G}}}

\def\gL{{\mathcal{L}}}

\def\gN{{\mathcal{N}}}

\def\gV{{\mathcal{V}}}

\usepackage{hyperref}
\usepackage{url}
\usepackage{graphicx}
\usepackage{enumitem}
\usepackage{subcaption}
\usepackage{algorithm}
\usepackage[noend]{algorithmic}

\title{Iterative Refinement Graph Neural Network for Antibody Sequence-Structure Co-design}

\author{Wengong Jin \\ Eric and Wendy Schmidt Center, Broad Institute of MIT and Harvard \\ \texttt{wengong@csail.mit.edu}
\AND
Jeremy Wohlwend, Regina Barzilay, Tommi Jaakkola  \\ CSAIL, Massachusetts Institute of Technology \\ \texttt{\{jwohlwend,regina,tommi\}@csail.mit.edu}
}

\iclrfinalcopy % Uncomment for camera-ready version, but NOT for submission.
\begin{document}

\maketitle
\begin{abstract}
Antibodies are versatile proteins that bind to pathogens like viruses and stimulate the adaptive immune system. The specificity of antibody binding is determined by complementarity-determining regions (CDRs) at the tips of these Y-shaped proteins. In this paper, we propose a generative model to automatically design the CDRs of antibodies with enhanced binding specificity or neutralization capabilities. Previous generative approaches formulate protein design as a structure-conditioned sequence generation task, assuming the desired 3D structure is given a priori. In contrast, we propose to co-design the sequence and 3D structure of CDRs as graphs. Our model unravels a sequence autoregressively while iteratively refining its predicted global structure. The inferred structure in turn guides subsequent residue choices. For efficiency, we model the conditional dependence between residues inside and outside of a CDR in a coarse-grained manner. Our method achieves superior log-likelihood on the test set and outperforms previous baselines in designing antibodies capable of neutralizing the SARS-CoV-2 virus.
\end{abstract}
\section{Introduction}
Monoclonal antibodies are increasingly adopted as therapeutics targeting a wide range of pathogens such as SARS-CoV-2~\citep{pinto2020cross}. Since the binding specificity of these Y-shaped proteins is largely determined by their complementarity-determining regions (CDRs), the main goal of computational antibody design is to automate the creation of CDR subsequences with desired properties. This problem is particularly challenging due to the combinatorial search space of over $20^{60}$ possible CDR sequences and the small solution space which satisfies the desired constraints of binding affinity, stability, and synthesizability \citep{raybould2019five}.

There are three key modeling questions in CDR generation. The first is how to model the relation between a sequence and its underlying 3D structure. Generating sequences without the corresponding structure~\citep{alley2019unified,shin2021protein} can lead to sub-optimal performance~\citep{ingraham2019generative}, while generating from a predefined 3D structure~\citep{ingraham2019generative} is not suitable for antibodies since the desired structure is rarely known a priori~\citep{fischman2018computational}.
Therefore, it is crucial to develop models that  \emph{co-design} the sequence and structure.
The second question is how to model the conditional distribution of CDRs given the remainder of a sequence (\emph{context}). Attention-based methods only model the conditional dependence at the sequence level, but the structural interaction between the CDR and its context is crucial for generation.
The last question relates to the model's ability to optimize for various properties. Traditional physics-based methods \citep{lapidoth2015abdesign,adolf2018rosettaantibodydesign} focus on binding energy minimization, but in practice, our objective can be much more involved than binding energies~\citep{liu2020antibody}.

In this paper, we represent a sequence-structure pair as a graph and formulate the co-design task as a graph generation problem.
The graph representation allows us to model the conditional dependence between a CDR and its context at both the sequence and structure levels.
Antibody graph generation poses unique challenges because the global structure is expected to change when new nodes are inserted. Previous autoregressive models \citep{you2018graphrnn,gebauer2019symmetry} cannot modify a generated structure because they are trained under teacher forcing. Thus errors made in the previous steps can lead to a cascade of errors in subsequent generation steps. 
To address these problems, we propose a novel architecture which interleaves the generation of amino acid nodes with the prediction of 3D structures. The structure generation is based on an iterative refinement of a global graph rather than a sequential expansion of a partial graph with teacher forcing.
Since the context sequence is long, we further introduce a coarsened graph representation by grouping nodes into blocks. We apply graph convolution at a coarser level to efficiently propagate the contextual information to the CDR residues.
After pretraining our model on antibodies with known structures, we finetune it using a predefined property predictor to generate antibodies with specific properties.

%Our model unravels a sequence autoregressively while revising its associated structure by repredicting pairwise distance and angles between all nodes.

%We then encode a revised graph structure using a message passing network to guide subsequent amino acid choices. Since our model only generates the CDR of an antibody with other parts fixed, we propose to coarsen the subgraph outside of a CDR for computational efficiency.

We evaluate our method on three generation tasks, ranging from language modeling to SARS-CoV-2 neutralization optimization and antigen-binding antibody design. Our method is compared with a standard sequence model~\citep{saka2021antibody,akbar2021silico} and a state-of-the-art graph generation method~\citep{you2018graphrnn} tailored to antibodies. Our method not only achieves lower perplexity on test sequences but also outperforms previous baselines in property-guided antibody design tasks.
\section{Related work}

%Protein/Antibody design (ABdesign, RAbD, Ge liu, Rosetta) 
%Generative models: Ingraham (+his related work, follow-ups), protein language models (need to incorporate more)...
\textbf{Antibody/protein design.}
Current methods for computational antibody design roughly fall into two categories. The first class is based on energy function optimization~\citep{pantazes2010optcdr,li2014optmaven,lapidoth2015abdesign,adolf2018rosettaantibodydesign}, which use Monte Carlo simulation to iteratively modify a sequence and its structure until reaching a local energy minimum. Similar approaches are used in protein design~\citep{leaver2011rosetta3,tischer2020design}.
Nevertheless, these physics-based methods are computationally expensive \citep{ingraham2019generative} and our desired objective can be much more complicated than low binding energy~\citep{liu2020antibody}.
%For example, it takes 30 seconds for Rosetta~\citep{leaver2011rosetta3} to design a sequence of 15 residues \citep{ingraham2019generative}.
%Similarly, \citet{liu2018constrained} trained a deep neural network to predict antibody binding affinity and used gradient ascent to optimize a discrete CDR sequence. However, their method do not model the 3D structure of antibodies.

The second class is based on generative models. For antibodies, they are mostly sequence-based \citep{alley2019unified,shin2021protein,saka2021antibody,akbar2021silico}. For proteins, \citet{o2018spin2,ingraham2019generative,strokach2020fast,karimi2020novo,cao2021fold2seq} further developed models conditioned on a backbone structure or protein fold.
Our model also seeks to incorporate 3D structure information for antibody generation. Since the best CDR structures are often unknown for new pathogens, we co-design sequences and structures for specific properties.

\textbf{Generative models for graphs.}
Our work is related to autoregressive models for graph generation \citep{you2018graphrnn,li2018learning,liu2018constrained,liao2019efficient,jin2020hierarchical}. In particular, \citet{gebauer2019symmetry} developed G-SchNet for molecular graph and conformation co-design. Unlike our method, they generate edges sequentially and cannot modify a previously generated subgraph when new nodes arrive. While Graphite~\citep{grover2019graphite} also uses iterative refinement to predict the adjacency matrix of a graph, it assumes all the node labels are given and predicts edges only. In contrast, our work combines autoregressive models with iterative refinement to generate a full graph with node and edge labels, including node labels and coordinates.

\textbf{3D structure prediction.}
Our approach is closely related to protein folding~\citep{ingraham2018learning,yang2020improved,baek2021accurate,jumper2021highly}. Inputs to the state-of-the-art models like AlphaFold require a complete protein sequence, its multi-sequence alignment (MSA), and its template features. These models are not directly applicable because we need to predict the structure of an \emph{incomplete} sequence and the MSA is not specified in advance.

Our iterative refinement model is also related to score matching methods for molecular conformation prediction \citep{shi2021learning} and diffusion-based methods for point clouds \citep{luo2021diffusion}. These algorithms also iteratively refine a predicted 3D structure, but only for a complete molecule or point cloud. In contrast, our approach learns to predict the 3D structure for incomplete graphs and interleaves 3D structure refinement with graph generation. 

\section{Antibody Sequence and Structure Co-Design}

\textbf{Overview.} The role of an antibody is to bind to an \textit{antigen} (e.g. a virus), present it to the immune system, and stimulate an immune response. A subset of antibodies known as \emph{neutralizing} antibodies not only bind to an antigen but can also suppress its activity. An antibody consists of a heavy chain and a light chain, each composed of one \textit{variable domain} (VH/VL) and some constant domains. The variable domain is further divided into a \textit{framework region} and three \textit{complementarity determining regions} (CDRs). The three CDRs on the heavy chain are labeled as CDR-H1, CDR-H2, CDR-H3, each occupying a \textit{contiguous} subsequence (Figure~\ref{fig:antibody}). 
As the most variable part of an antibody, CDRs are the main determinants of binding and neutralization~\citep{abbas2014cellular}.

Following \citet{shin2021protein,akbar2021silico}, we formulate antibody design as a CDR generation task, conditioned on the framework region. 
Specifically, we represent an antibody as a graph, which encodes both its sequence and 3D structure. We propose a new graph generation approach called RefineGNN and extend it to handle conditional generation given a fixed framework region. Lastly, we describe how to apply RefineGNN to property-guided optimization to design new antibodies with better neutralization properties.
For simplicity, we focus on the generation of heavy chain CDRs, though our method can be easily extended to model light chains CDRs. 

\textbf{Notations.} An antibody VH domain is represented as a sequence of amino acids $\vs = \vs_1\vs_2\cdots\vs_n$. Each token $\vs_i$ in the sequence is called a \emph{residue}, whose value can be either one of the 20 amino acids or a special token $\pad$, meaning that its amino acid type is unknown and needs to be predicted. 
The VH sequence folds into a 3D structure and each residue $\vs_i$ is labeled with three backbone coordinates: $\vx_{i,\alpha}$ for its alpha carbon atom, $\vx_{i,c}$ for its carbon atom, and $\vx_{i,n}$ for its nitrogen atom.

\begin{figure}[t]
    \centering
    \includegraphics[width=\textwidth]{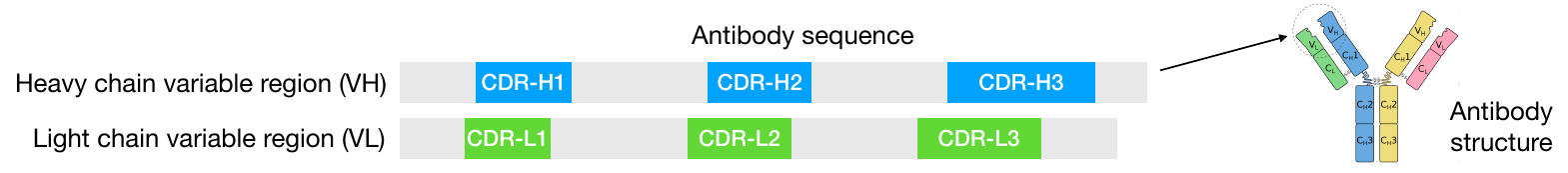}
    \caption{Schematic structure of an antibody (figure modified from Wikipedia).}
    \vspace{-5pt}
    \label{fig:antibody}
\end{figure}

\subsection{Graph representation} 
We represent an antibody (VH) as a graph $\gG(\vs)=(\gV,\gE)$ with node features $\gV = \{\vv_1, \cdots, \vv_n\}$ and edge features $\gE = \{\ve_{ij}\}_{i \neq j}$. Each node feature $\vv_i$ encodes three dihedral angles $(\phi_i, \psi_i, \omega_i)$ related to three backbone coordinates of residue $i$. 
For each residue $i$, we compute an orientation matrix $\mO_i$ representing its local coordinate \emph{frame} \citep{ingraham2019generative} (defined in the appendix). This allows us to compute edge features describing the spatial relationship between two residues $i$ and $j$ (i.e. distance and orientation) as follows
\begin{equation}
    \ve_{ij} = \big(
    E_{\mathrm{pos}}(i - j), \quad
    \mathrm{RBF}(\Vert \vx_{i,\alpha} - \vx_{j,\alpha} \Vert),\quad
    \mO_i^\top \frac{\vx_{j,\alpha} - \vx_{i,\alpha}}{\Vert \vx_{i,,\alpha} - \vx_{j,,\alpha} \Vert},\quad
    \vq(\mO_i^\top \mO_j)
    \big). \label{eq:edge-feature}
\end{equation}
The edge feature $\ve_{ij}$ contains four parts. The positional encoding $E_{\mathrm{pos}}(i - j)$ encodes the relative distance between two residues in an antibody sequence. 
The second term $\mathrm{RBF}(\cdot)$ is a \emph{distance} encoding lifted into radial basis. 
The third term in $\ve_{ij}$ is a \emph{direction} encoding that corresponds to the relative direction of $\vx_j$ in the local frame of residue $i$. 
The last term $\vq(\mO_i^\top \mO_j)$ is the \emph{orientation} encoding of the quaternion representation $\vq(\cdot)$ of the spatial rotation matrix $\mO_i^\top \mO_j$.
We only include edges in the $K$-nearest neighbors graph of $\gG(\vs)$ with $K=8$.
For notation convenience, we use $\gG$ as a shorthand for $\gG(\vs)$ when there is no ambiguity.

\subsection{Iterative Refinement Graph Neural Network (RefineGNN)}

\begin{figure}[t]
    \centering
    \includegraphics[width=\textwidth]{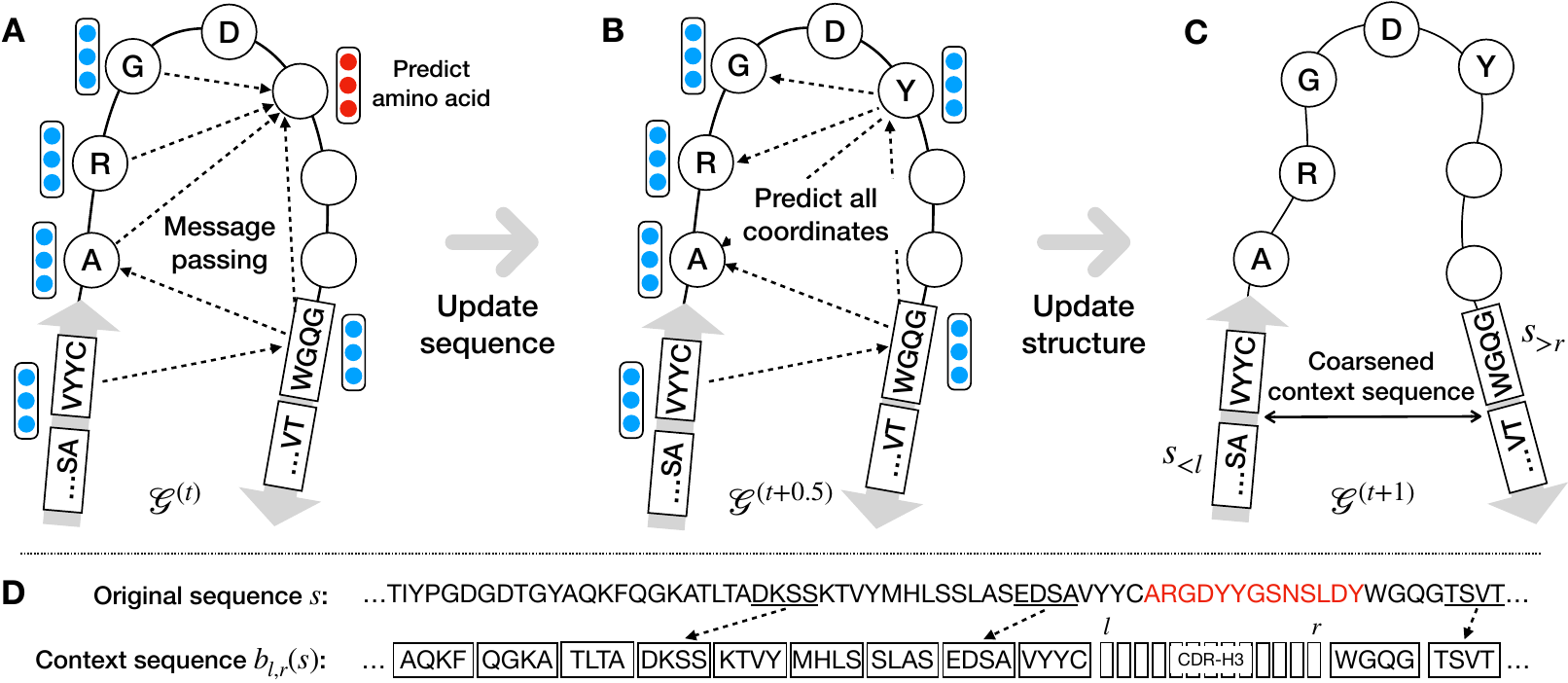}
    \caption{(A-C) One generation step of RefineGNN. Each circle represents a CDR residue and each square represents a residue block in a coarsened context sequence. (D) Sequence coarsening.}
    \label{fig:model}
\end{figure}

We propose to generate an antibody graph via an iterative refinement process. Let $\gG^{(0)}$ be the initial guess of the true antibody graph. Each residue is initialized as a special token $\pad$ and each edge $(i,j)$ is initialized to be of distance $3|i-j|$ since the average distance between consecutive residues is around three. The direction and orientation features are set to zero.
In each generation step $t$, the model learns to revise a current antibody graph $\gG^{(t)}$ and predict the label of the next residue $t+1$. 
Specifically, it first encodes $\gG^{(t)}$ with a message passing network (MPN) with parameter $\theta$ 
\begin{equation}
    \set{\vh_1^{(t)},\cdots,\vh_n^{(t)}} = \MPN_{\theta}(\gG^{(t)}),
\end{equation}
where $\vh_i^{(t)}$ is a learned representation of residue $i$ under the current graph $\gG^{(t)}$. Our MPN consists of $L$ message passing layers with the following architecture 
\begin{equation}
    \vh_i^{(t, l+1)} = \mathrm{LayerNorm}\bigg(\sum_j \FFN\big(\vh_i^{(t, l)}, \vh_j^{(t, l)}, E(\vs_j), \ve_{i,j}\big)\bigg), \quad 0 \leq l \leq L-1,
\end{equation}
where $\vh_i^{(t,0)} = \vv_i$ and $\vh_i^{(t)} = \vh_i^{(t, L)}$. $\FFN$ is a two-layer feed-forward network (FFN) with ReLU activation function. $E(\vs_j)$ is a learned embedding of amino acid type $\vs_j$.
Based on the learned residue representations, we predict the amino acid type of the next residue $t+1$ (Figure~\ref{fig:model}A).
\begin{equation}
    \vp_{t+1} = \mathrm{softmax}(\mW_a \vh_{t+1}^{(t)}) \label{eq:aa-pred}
\end{equation}
This prediction gives us a new graph $\gG^{(t+0.5)}$ with the same edges as $\gG^{(t)}$ but the node label of $t+1$ is changed (Figure~\ref{fig:model}B). Next, we need to update the structure to accommodate the new residue $t+1$. To this end, we encode graph $\gG^{(t+0.5)}$ by another MPN with a different parameter $\tilde{\theta}$ and predict the coordinate of all residues.
\begin{eqnarray}
    \set{\vh_1^{(t+0.5)},\cdots,\vh_n^{(t+0.5)}} &=& \MPN_{\tilde{\theta}}(\gG^{(t+0.5)}) \\
    \vx_{i,e}^{(t+1)} &=& \mW_x^e \vh_i^{(t+0.5)}, \qquad 1 \leq i \leq n, e \in \set{\alpha, c, n}. \label{eq:coord-pred}
\end{eqnarray}
The new coordinates $\vx_{i}^{(t+1)}$ define a new antibody graph $\gG^{(t+1)}$ for the next iteration (Figure~\ref{fig:model}C). We explicitly realize the coordinates of each residue because we need to calculate the spatial edge features for $\gG^{(t+1)}$.
The structure prediction (coordinates $\vx_i$) and sequence prediction (amino acid types $\vp_{t+1}$) are carried out by two different MPNs, namely the \textit{structure} network $\tilde{\theta}$ and \textit{sequence} network $\theta$. This disentanglement allows the two networks to focus on two distinct tasks.

\textbf{Training.} During training, we only apply teacher forcing to the discrete amino acid type prediction. Specifically, in each generation step $t$, residues $1$ to $t$ are set to their ground truth amino acid types $\vs_1,\cdots, \vs_t$, while all future residues $t+1, \cdots, n$ are set to a padding token. In contrast, the continuous structure prediction is carried out \emph{without} teacher forcing. In each iteration, the model refines the \emph{entire} structure predicted in the previous step and constructs a new $K$-nearest neighbors graph $\gG^{(t+1)}$ of \emph{all} residues based on the predicted coordinates $\set{\vx_{i,e}^{(t+1)} \;|\; 1 \leq i \leq n, e\in \set{\alpha,c,n}}$.

\textbf{Loss function.} Our model remains rotation and translation invariant because the loss function is computed over pairwise distance and angles rather than coordinates. The loss function for antibody structure prediction consists of three parts.
\begin{itemize}[leftmargin=*,topsep=0pt,itemsep=0pt]
    \item \emph{Distance loss}: For each residue pair $i, j$, we compute its pairwise distance between the predicted alpha carbons $\vx_{i,\alpha}^{(t)}, \vx_{j,\alpha}^{(t)}$. We define the distance loss as the Huber loss between the predicted and true pairwise distances
    \begin{equation}
        \mathcal{L}_d^{(t)} = \sum_{i,j}\nolimits \ell_{\mathrm{huber}}(\Vert \vx_{i,\alpha}^{(t)} - \vx_{j,\alpha}^{(t)} \Vert^2, \Vert \vx_{i,\alpha} - \vx_{j,\alpha} \Vert^2),
    \end{equation}
    where distance is squared to avoid the square root operation which causes numerical instability.
    
    \item \emph{Dihedral angle loss}: For each residue, we calculate its dihedral angle $\big(\phi_i^{(t)}, \psi_i^{(t)}, \omega_i^{(t)}\big)$ based on the predicted atom coordinates $\vx_{i,\alpha}^{(t)}, \vx_{i,c}^{(t)}, \vx_{i,n}^{(t)}$ and $\vx_{i+1,\alpha}^{(t)}, \vx_{i+1,c}^{(t)}, \vx_{i+1,n}^{(t)}$. We define the dihedral angle loss as the mean square error between the predicted and true dihedral angles
    \begin{equation}
        \mathcal{L}_a^{(t)} = \sum_{i} \sum_{a \in \set{\phi,\psi,\omega}}(\cos a_i^{(t)} - \cos a_i)^2 + (\sin a_i^{(t)} - \sin a_i)^2
    \end{equation}
    
    \item \emph{$C_\alpha$ angle loss}: We calculate angles $\gamma_i^{(t)}$ between two vectors $\vx_{i-1,\alpha}^{(t)} - \vx_{i,\alpha}^{(t)}$ and $\vx_{i,\alpha}^{(t)} - \vx_{i+1,\alpha}^{(t)}$ as well as dihedral angles $\beta_i^{(t)}$ between two planes defined by $\vx_{i-2,\alpha}^{(t)}, \vx_{i-1,\alpha}^{(t)}, \vx_{i,\alpha}^{(t)}, \vx_{i+1,\alpha}^{(t)}$.
    \begin{equation}
        \mathcal{L}_c^{(t)} = \sum_{i}\nolimits (\cos\gamma_i^{(t)} - \cos\gamma_i)^2 + (\cos\beta_i^{(t)} - \cos\beta_i)^2
    \end{equation}
\end{itemize}
In summary, the overall graph generation loss is defined as $\mathcal{L} = \mathcal{L}_\mathrm{seq} + \mathcal{L}_\mathrm{struct}$, where
\begin{equation}
    \mathcal{L}_\mathrm{struct} = \sum_t\nolimits \mathcal{L}_d^{(t)} + \mathcal{L}_a^{(t)} + \mathcal{L}_c^{(t)} \qquad \mathcal{L}_\mathrm{seq} = \sum_t\nolimits \gL_{ce}(\vp_t, \vs_t).  \label{eq:objective}
\end{equation}
The sequence prediction loss $\mathcal{L}_\mathrm{seq}$ is the cross entropy $\gL_{ce}$ between predicted and true residue types.

\subsection{Conditional generation given the framework region} \label{sec:conditional}

The model architecture described so far is designed for unconditional generation --- it generates an entire antibody graph without any constraints. In practice, we usually fix the framework region of an antibody and design the CDR sequence only. Therefore, we need to extend the model architecture to learn the conditional distribution $P(\vs' | \vs_{<l}, \vs_{>r})$, where $\vs_{<l}=\vs_1\cdots\vs_{l-1}$ and $\vs_{>r}=\vs_{r+1}\cdots\vs_n$ are residues outside of the CDR $\vs_l \cdots \vs_r$.

\textbf{Conditioning via attention.} A simple extension of RefineGNN is to encode the non-CDR sequence using a recurrent neural network and propagate information to the CDR through an attention layer. To be specific, we first concatenate $\vs_{<l}$ and $\vs_{>r}$ into a \emph{context} sequence $\tilde{\vs} = \vs_{<l} \oplus \pad \cdots \pad \oplus \vs_{>r}$, where $\oplus$ means string concatenation and $\pad$ is repeated $n$ times. We then encode this context sequence by a Gated Recurrent Unit (GRU) \citep{cho2014learning} and modify the structure and sequence prediction step (Equation \ref{eq:aa-pred} and \ref{eq:coord-pred}) as
\begin{eqnarray}
    \set{\vc_1, \cdots, \vc_n} &=& \vc_{1:n} = \mathrm{GRU}(\tilde{\vs}) \\
    \vp_{t+1} &=& \mathrm{softmax}\big(\mW_a \vh_{t+1}^{(t)} + \mU_a^\top \mathrm{attention}(\vc_{1:n}, \vh_{t+1}^{(t)})\big) \\
    \vx_{i,e}^{(t+1)} &=& \mW_x^e \vh_i^{(t+0.5)} + \mU_x^e{}^\top \mathrm{attention}(\vc_{1:n}, \vh_{i}^{(t+0.5)}) \label{eq:coarse}
\end{eqnarray}

\textbf{Multi-resolution modeling.} The attention-based approach alone is not sufficient because it does not model the structure of the context sequence, thus ignoring how its residues \emph{structurally} interact with the CDR's. While this information is not available for new antibodies at test time, we can learn to predict this interaction using antibodies in the training set with known structures.

A naive solution is to iteratively refine the entire antibody structure (more than 100 residues) while generating CDR residues. This approach is computationally expensive because we need to recompute the MPN encoding for all residues in each generation step. Importantly, we cannot predict the context residue coordinates at the outset and fix them because they need to be adjusted accordingly when the coordinates of CDR residues are updated in each generation step.

For computational efficiency, we propose a coarse-grained model that reduces the context sequence length by clustering it into \emph{residue blocks}. Specifically, we construct a coarsened context sequence $\vb_{l,r}(\vs)$ by clustering every $b$ context residues into a block (Figure~\ref{fig:model}D).
The new sequence $\vb_{l,r}(\vs)$ defines a coarsened graph $\gG(\vb_{l,r}(\vs))$ over the residue blocks, whose edges are defined based on block coordinates. 
The coordinate of each block $\vx_{\vb_i,e}$ is defined as the mean coordinate of residues within the block. The embedding of each block $E(\vb_i)$ is the mean of its residue embeddings.
\begin{equation}
E(\vb_i) = \sum_{\vs_j \in \vb_i}\nolimits E(\vs_j) / b, \qquad \vx_{\vb_i,e} = \sum_{\vs_j \in \vb_i}\nolimits \vx_{j,e} / b, \qquad e \in \set{\alpha, c, n}.
\end{equation} 
Now we can apply RefineGNN to generate the CDR residues while iteratively refining the global graph $\gG(\vb_{l,r}(\vs))$ by predicting the coordinates of all blocks. The only change is that the structure prediction loss is defined over block coordinates $\vx_{\vb_i,e}$. Lastly, we combine both the attention mechanism and coarse-grained modeling to keep both fine-grained and coarse-grained information. The decoding process of this conditional RefineGNN is illustrated in Algorithm~\ref{alg:decode}.

\subsection{Property-guided sequence optimization} \label{sec:finetune}
Our ultimate goal is to generate new antibodies with desired properties such as neutralizing a particular virus. This task can be formulated as an optimization problem. Let $Y$ be a binary indicator variable for neutralization. Our goal is to learn a conditional generative model $P_\Theta(\vs' | \vb_{l,r}(\vs))$ that maximizes the probability of neutralization for a training set of antibodies $\gD$, i.e. 
\begin{equation}
     \sum_{\vs \in \gD}\nolimits \log P(Y=1 | \vb_{l,r}(\vs)) = \sum_{\vs \in \gD}\nolimits \log \sum_{\vs'}\nolimits f(\vs') P_\Theta(\vs' | \vb_{l,r}(\vs)) 
\end{equation}
where $f(\vs')$ is a predictor for $P(Y=1 | \vs')$. Assuming $f$ is given, this problem can be solved by iterative target augmentation (ITA) \citep{yang2020improving}. Before ITA optimization starts, we first pretrain our model on a set of real antibody structures to learn a prior distribution over CDR sequences and structures. In each ITA finetuning step, we first randomly sample a sequence $\vs$ from $\gD$, a set of antibodies whose CDRs need to be redesigned. Next, we generate $M$ new sequences given its context $\vb_{l,r}(\vs)$. A generated sequence $\vs_i'$ is added to our training set $Q$ if it is predicted as neutralizing. Initially, the training set $Q$ contains antibodies that are known to be neutralizing ($Y=1$). Lastly, we sample a batch of neutralizing antibodies from $Q$ and update the model parameter by minimizing their sequence prediction loss $\mathcal{L}_\mathrm{seq}$ (Eq.(\ref{eq:objective})). The structure prediction loss $\mathcal{L}_\mathrm{struct}$ is excluded in ITA finetuning phase because the structure of a generated sequence is unknown. 

\begin{figure}[t]
\begin{minipage}{0.48\textwidth}
    \begin{algorithm}[H]
    \begin{algorithmic}[1]
    \caption{RefineGNN decoding}
    \label{alg:decode}
    \REQUIRE Context sequence $\vs_{<l}, \vs_{>r}$
    \STATE Predict the CDR length $n$
    \STATE Coarsen the context sequence into $\vb_{l,r}(\vs)$
    \STATE Construct the initial graph $\gG^{(0)}$
    \FOR{$t = 0 \text{ to } n-1$}
    \STATE Encode $\gG^{(t)}$ using the sequence MPN
    \STATE Predict distribution of the next residue $\vp_{t+1}$
    \STATE Sample $\vs_{t+1} \sim \mathrm{categorical}(\vp_{t+1})$
    \STATE Encode $\gG^{(t+0.5)}$ with the structure MPN
    \STATE Predict all residue coordinates $\vx_{i,e}^{(t+1)}$
    \STATE Update $\gG^{(t+1)}$ using the new coordinates
    \ENDFOR
    \end{algorithmic}
    \end{algorithm}
\end{minipage}
~
\begin{minipage}{0.50\textwidth}
    \begin{algorithm}[H]
    \begin{algorithmic}[1]
    \caption{ITA-based sequence optimization}
    \label{alg:emopt}
    \REQUIRE A set of antibodies $\gD$ to be optimized
    \REQUIRE A neutralization predictor $f$.
    \REQUIRE A set of neutralizing antibodies $Q$
    \FOR{each iteration}
    \STATE Sample an antibody $\vs$ from $\gD$, remove its CDR and get a context sequence $\vb_{l,r}(\vs)$
    \FOR{$i = 1 \text{ to } M$}
        \STATE Sample $\vs_i' \sim P_\Theta(\vs' | \vb_{l,r}(\vs))$
        \IF{$f(\vs_i') > \max(f(\vs), 0.5)$}
        \STATE $Q \leftarrow Q \cup \set{\vs_i'}$
        \ENDIF
    \ENDFOR
    \STATE Sample a batch of new antibodies from $Q$
    \STATE Update model parameter $\Theta$ by minimizing the sequence prediction loss $\mathcal{L}_\mathrm{seq}$.
    \ENDFOR
    \end{algorithmic}
    \end{algorithm}
\end{minipage}
\end{figure}
\section{Experiments}
\textbf{Setup.} We construct three evaluation setups to quantify the performance of our approach. Following standard practice in generative model evaluation, we first measure the perplexity of different models on new antibodies in a test set created based on sequence similarity split.
%Indeed, a good generative model should not only achieve a low perplexity, but also a low structure prediction error on the test set. 
We also measure structure prediction error by comparing generated and ground truth CDR structures recorded in the Structural Antibody Database \citep{dunbar2014sabdab}. 
Results for this task are shown in section~\ref{sec:lm}.

Second, we evaluate our method on an existing antibody design benchmark of 60 antibody-antigen complexes from \citet{adolf2018rosettaantibodydesign}. The goal is to design the CDR-H3 of an antibody so that it binds to a given antigen. Results for this task are shown in section~\ref{sec:rabd}.

Lastly, we propose an antibody optimization task which aims to redesign CDR-H3 of antibodies in the Coronavirus Antibody Database~\citep{raybould2021cov} to improve their neutralization against SARS-CoV-2. CDR-H3 design with a fixed framework is a common practice in the antibody engineering community~\citep{adolf2018rosettaantibodydesign,liu2020antibody}. Following works in molecular design~\citep{jin2020multi}, we use a predictor to evaluate the neutralization of generated antibodies since we cannot experimentally test them in wet labs.
Results for this task are reported in section~\ref{sec:sars2}.

\textbf{Baselines.} We consider three baselines for comparison (details in the appendix). The first baseline is a sequence-based LSTM model used in \cite{saka2021antibody,akbar2021silico}. This model does not utilize any 3D structure information. It consists of an encoder that learns to encode a context sequence $\tilde{\vs}$, a decoder that decodes a CDR sequence, and an attention layer connecting the two.

The second baseline is an autoregressive graph generation model (AR-GNN) whose architecture is similar to \citet{you2018graphrnn,jin2020multi} but tailored for antibodies. AR-GNN generates an antibody graph residue by residue. In each step $t$, it first predicts the amino acid type of residue $t$ and then generates edges between $t$ and previous residues.
Importantly, AR-GNN cannot modify a partially generated 3D structure of residues $\vs_1 \cdots \vs_{t-1}$ because it is trained by teacher forcing.
%AR-GNN is different from RefineGNN because it cannot revise a previously generated structure (pairwise distance ).

On the antigen-binding task, we include an additional physics-based baseline called RosettaAntibodyDesign (RAbD)~\citep{adolf2018rosettaantibodydesign}. We apply their \textit{de novo} design protocol composed of graft design followed by 250 iterations of sequence design and energy minimization. We cannot afford to run more iterations because it takes more than 10 hours per antibody. We also could not apply RAbD to the SARS-CoV-2 task because it requires 3D structures to be given. This information is unavailable for antibodies in CoVAbDab.

\textbf{Hyperparameters.} We performed hyperparameter tuning to find the best setting for each method. For RefineGNN, both its structure and sequence MPN have four message passing layers, with a hidden dimension of 256 and block size $b=4$. All models are trained by the Adam optimizer with a learning rate of 0.001. More details are provided in the appendix.

\begin{table}[t]
\centering
\caption{\emph{Left}: Language modeling results. We report perplexity (PPL) and root mean square deviation (RMSD) for each CDR in the heavy chain. \emph{Right}: Results on the antigen-binding antibody design task. We report the amino acid recovery (AAR) for all methods.}
\vspace{-4pt}
\label{tab:stat}
\begin{subtable}{.7\textwidth}
\centering
\begin{tabular}{lcccccc}
\hline
& \multicolumn{2}{c}{CDR-H1} & \multicolumn{2}{c}{CDR-H2}  & \multicolumn{2}{c}{CDR-H3}  \Tstrut\Bstrut \\
\cline{2-7}
Model & PPL & RMSD & PPL & RMSD & PPL & RMSD  \Tstrut\Bstrut \\
\hline
LSTM & 6.79 & - & 7.21 & - & 9.70 & - \Tstrut\Bstrut\\
AR-GNN & 6.44 & 2.97 & 6.86 & 2.27 & 9.44 & 3.63 \Tstrut\Bstrut\\
\hline
RefineGNN & \textbf{6.09} & \textbf{1.18} & \textbf{6.58} & \textbf{0.87} & \textbf{8.38} & \textbf{2.50} \Tstrut\Bstrut \\
\hline
\end{tabular}
\end{subtable}
~
\begin{subtable}{.28\textwidth}
\centering
\begin{tabular}{lc}
\hline
Model & AAR  \Tstrut\Bstrut \\
\hline
RAbD & 28.53\% \Tstrut\Bstrut\\
\hline
LSTM & 22.53\%  \Tstrut\Bstrut\\
AR-GNN & 23.86\%  \Tstrut\Bstrut\\
\hline
RefineGNN & \textbf{35.37\%} \Tstrut\Bstrut \\
\hline
\end{tabular}
\end{subtable}
\vspace{-5pt}
\end{table}

\subsection{Language modeling and 3D structure prediction} \label{sec:lm}

\textbf{Data.} The Structural Antibody Database (SAbDab) consists of 4994 antibody structures renumbered according to the IMGT numbering scheme \citep{lefranc2003imgt}. %Each structure is recorded in the Protein Data Bank (PDB) format and contains the heavy and light chains of an antibody and chains of an antigen. %In total, there are 8795 redundant heavy chains and 7104 redundant light chains.
To measure a model's ability to generalize to novel CDR sequences, we divide the heavy chains into training, validation, and test sets based on CDR cluster split. We illustrate our cluster split process using CDR-H3 as an example. First, we use MMseqs2 \citep{steinegger2017mmseqs2} to cluster all the CDR-H3 sequences. The sequence identity is calculated under the BLOSUM62 substitution matrix \citep{henikoff1992amino}. Two antibodies are put into the same cluster if their CDR-H3 sequence identity is above 40\%. We then randomly split the clusters into training, validation, and test set with 8:1:1 ratio. We repeat the same procedure for creating CDR-H1 and CDR-H2 splits. In total, there are 1266, 1564, and 2325 clusters for CDR-H1, H2, and H3. The size of training, validation, and test sets for each CDR is shown in the appendix.

\textbf{Metrics.} For each method, we report the perplexity (PPL) of test sequences and the root mean square deviation (RMSD) between a predicted structure and its ground truth structure reported in SAbDab. RMSD is calculated by the Kabsch algorithm~\citep{kabsch1976solution} based on $C_\alpha$ coordinate of CDR residues. Since the mapping between sequences and structures is deterministic in RefineGNN, we can calculate perplexity in the same way as standard sequence models.

\begin{figure}[t]
    \centering
    \includegraphics[width=0.9\textwidth]{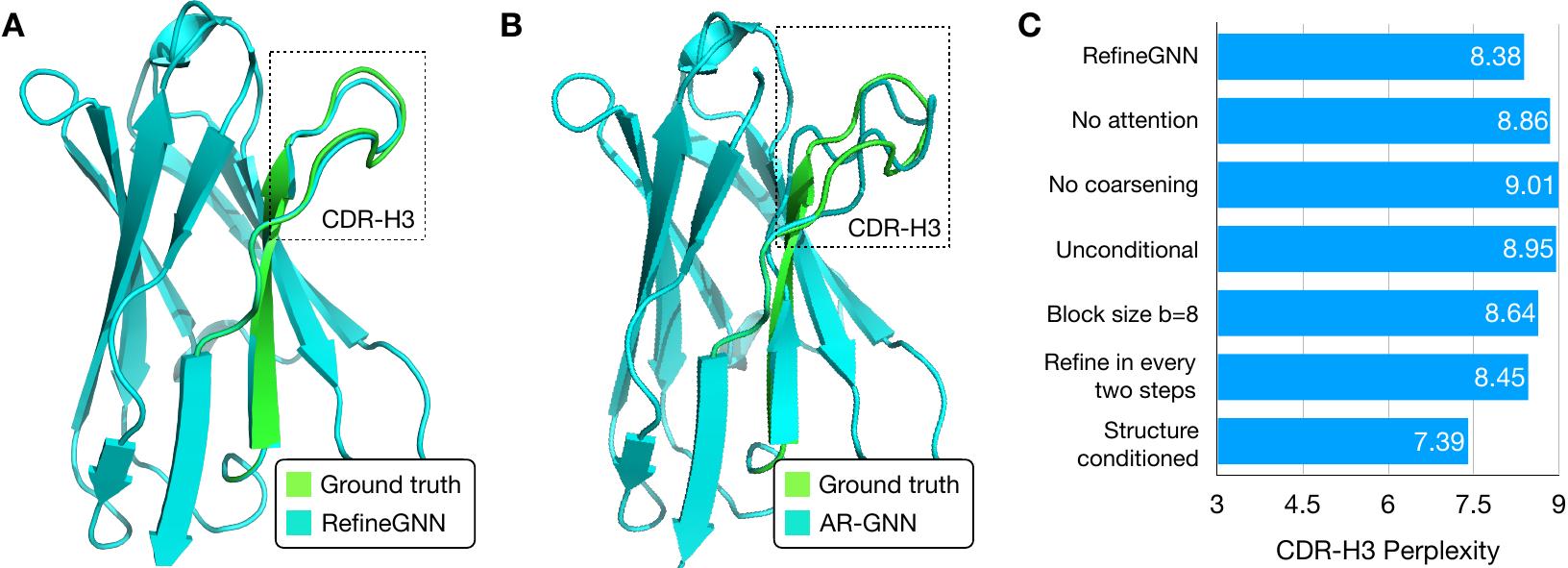}
    \caption{(A) CDR-H3 structure predicted by RefineGNN (PDB: 4bkl, RMSD = 0.57). The predicted structure (cyan) is aligned to the true structure (green) using the Kabsch algorithm. (B) CDR-H3 structure predicted by AR-GNN (PDB: 4bkl, RMSD = 2.16). (C) Ablation studies of different modeling choices in RefineGNN in the CDR-H3 perplexity evaluation task.}
    \label{fig:proteins}
    \vspace{-5pt}
\end{figure}

\textbf{Results.} Since the LSTM baseline does not involve structure prediction, we report RMSD for graph-based methods only. As shown in Table~\ref{tab:stat}, RefineGNN significantly outperforms all baselines in both metrics. For CDR-H3, our model gives 13\% PPL reduction (8.38 v.s.~9.70) over sequence only model and 10\% PPL reduction over AR-GNN (8.38 v.s.~9.44). RefineGNN also predicts the structure more accurately, with 30\% relative RMSD reduction over AR-GNN. In Figure~\ref{fig:proteins}, we provide examples of predicted 3D structures of CDR-H3 loops.

\textbf{Ablation studies.} We further conduct ablation experiments on the CDR-H3 generation task to study the importance of different modeling choices. First, when we remove the attention mechanism and context coarsening step in section \ref{sec:conditional}, the PPL increases from 8.38 to 8.86 (Figure \ref{fig:proteins}C, row 2) and 9.01 (Figure \ref{fig:proteins}C, row 3) respectively. We also tried to remove both the attention and coarsening modules and trained the model without conditioning on the context sequence. The PPL of this unconditional variant is much worse than our conditional model (Figure \ref{fig:proteins}C, row 4). In short, these results validate the advantage of our multi-resolution conditioning strategy.

Moreover, the model performance slightly degrades when we halve the number of refinement steps or increase block size to $b=8$ (Figure \ref{fig:proteins}C, row 5-6) . Lastly, we train a structure-conditioned model by feeding the ground truth structure to RefineGNN at every generation step (Figure \ref{fig:proteins}C, row 7). While this structure-conditioned model gives a lower PPL as expected (7.39 v.s.~8.38), it is not too far away from the sequence only model (PPL = 9.70). This suggests that RefineGNN is able to extract a decent amount of information from the partial structure co-evolving with the sequence.

\subsection{Antigen-binding antibody design} \label{sec:rabd}

\textbf{Data.} \citet{adolf2018rosettaantibodydesign} selected 60 antibody-antigen complexes as an antibody design benchmark. Given the framework of an antibody, the goal is to design its CDR-H3 that binds to its corresponding antigen. For simplicity, none of the methods is conditioned on the antigen structure during CDR-H3 generation. We leave antigen-conditioned CDR generation for future work.

\textbf{Metric.} Following \citet{adolf2018rosettaantibodydesign}, we use amino acid recovery (AAR) as the evaluation metric. For any generated sequence, we define its AAR as the percentage of residues having the same amino acid as the corresponding residue in the original antibody.

\begin{figure}[t]
    \centering
    \includegraphics[width=\textwidth]{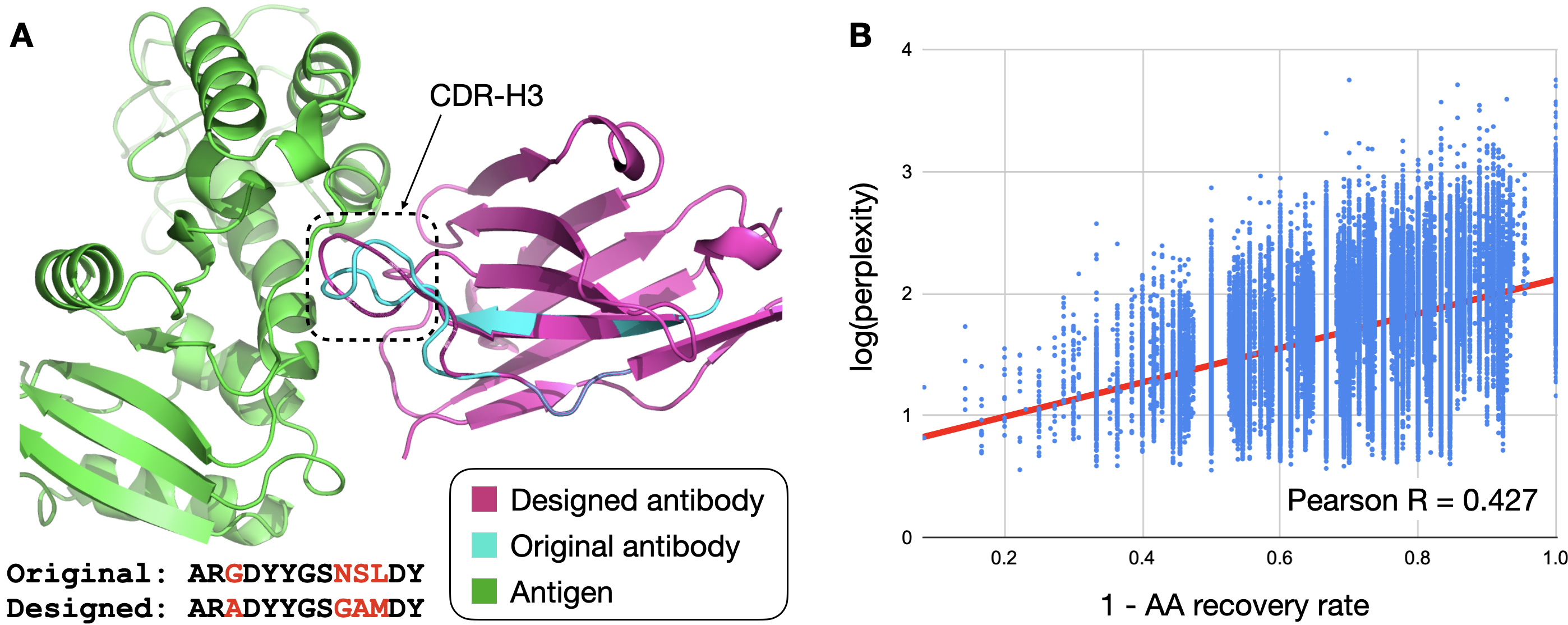}
    \caption{(A) Visualization of a generated CDR-H3 sequence and its structure in complex with an antigen (PDB: 4cmh). The predicted structure is aligned and grafted onto the original antibody using the Kabsch algorithm. Residues different from the original antibody are highlighted in red. (B) The correlation between the perplexity of a generated sequence and AA recovery error.}
    \label{fig:complex}
\end{figure}

\textbf{Results.} For LSTM, AR-GNN, and RefineGNN, the training set in this setup is the entire SAbDab except antibodies in the same cluster as any of the test antibodies. At test time, we generate 10000 CDR-H3 sequences for each antibody and select the top 100 candidates with the lowest perplexity. For simplicity, all methods are configured to generate CDRs of the same length as the original CDR. As shown in Table~\ref{tab:stat}, our model achieves the highest AAR score, with around 7\% absolute improvement over the best baseline.
In Figure~\ref{fig:complex}A, we show an example of a generated CDR-H3 sequence and highlight residues that are different from the original antibody. We also found that sequences with lower perplexity tend to have a lower AA recovery error (Pearson R = 0.427, Figure~\ref{fig:complex}B). This suggests that we can use perplexity as the ranking criterion for antibody design.

\subsection{SARS-CoV-2 neutralization optimization} \label{sec:sars2}

\textbf{Data.} The Coronavirus Antibody Database (CoVAbDab) contains 2411 antibodies, each associated with multiple binary labels indicating whether it neutralizes a coronavirus (SARS-CoV-1 or SARS-CoV-2) at a certain epitope. Similar to the previous experiment, we divide the antibodies into training, validation, and test sets based on CDR-H3 cluster split with 8:1:1 ratio.

\textbf{Neutralization predictor.} The predictor takes as input the VH sequence of an antibody and outputs a neutralization probability for the SARS-CoV-1 and SARS-CoV-2 viruses. Each residue is embedded into a 64 dimensional vector, which is fed to a SRU encoder~\citep{lei2021attention} followed by average-pooling and a two-layer feed forward network. The final outputs are the probabilities $p_1$ and $p_2$ of neutralizing SARS-CoV-1 and SARS-CoV-2 and our scoring function is $f(\vs)=p_2$. The predictor achieved 0.81 test AUROC for SARS-CoV-2 neutralization prediction.

\textbf{CDR sequence constraints.} Therapeutic antibodies must be free from developability issues such as glycosylation and high charges~\citep{raybould2019five}. Thus, we include four constraints on a CDR-H3 sequence $\vs$: 
1) Its net charge must be between -2.0 and 2.0~\citep{raybould2019five}. The definition of net charge is given in the appendix.
2) It must not contain the N-X-S/T motif which is prone to glycosylation.
3) Any amino acid should not repeat more than five times (e.g. SSSSS). 4) Perplexity of a generated sequence given by LSTM, AR-GNN, and RefineGNN should be all less than 10. The last two constraints force generated sequences to be realistic. We use all three models in the perplexity constraint to ensure a fair comparison for all methods.

\textbf{Metric.} For each antibody in the test set, we generate 100 new CDR-H3 sequences, concatenate them with its context sequence to form 100 full VH sequences, and feed them into the neutralization predictor $f$. We report the average neutralization score of antibodies in the test set. Neutralization score of a generated sequence $\vs'$ equals $f(\vs')$ if it satisfies all the CDR sequence constraints. Otherwise the score is the same as the original sequence. In addition, we pretrain each model on the SAbDab CDR-H3 sequences and evaluate its PPL on the CoVAbDab CDR-H3 sequences.

\textbf{Results.} All methods are pretrained on SAbDab antibodies and finetuned on CoVAbDab using the ITA algorithm to generate neutralizing antibodies. Our model outperforms the best baseline by a 3\% increase in terms of average neutralization score (Table~\ref{tab:covid}). Our pretrained RefineGNN also achieves a much lower perplexity on CoVAbDab antibodies (7.86 v.s.~8.67). 
Examples of generated CDR-H3 sequences and their predicted neutralization scores are shown in the appendix.

\begin{table}[t]
\centering
\caption{SARS-CoV-2 neutralization optimization results. For each method, we report the PPL on CoVAbDab after pretraining on SAbDab and then report the average neutralization score after ITA finetuning. The average neutralization probability of original CoVAbDab antibodies is 69.3\%. }
\vspace{-4pt}
\label{tab:covid}
\begin{tabular}{lcccc}
\hline
     & Original & LSTM & AR-GNN & RefineGNN \Tstrut\Bstrut \\
    \hline
    CoVAbDab PPL ($\downarrow$) & - & 9.40 & 8.67 & \textbf{7.86} \Tstrut\Bstrut \\
    Neutralization ($\uparrow$)& 69.3\% & 72.0\% & 70.4\% & \textbf{75.2\%} \Tstrut\Bstrut \\
\hline
\end{tabular}
\vspace{-5pt}
\end{table}

\section{Conclusion}
In this paper, we developed a RefineGNN model for antibody sequence and structure co-design. The advantage of our model over previous graph generation methods is its ability to revise a generated subgraph to accommodate addition of new residues. Our approach significantly outperforms sequence-based and graph-based approaches on three antibody generation tasks.

\subsection*{Acknowledgement}
We would like to thank Rachel Wu, Xiang Fu, Jason Yim, and Peter Mikhael for their valuable feedback on the manuscript. We also want to thank Nitan Shalon, Nicholas Webb, Jae Hyeon Lee, Qiu Yu, and Galit Alter for their suggestions on method development. We are grateful for the generous support of Mark and Lisa Schwartz, funding in a form of research grant from Sanofi, Defense Threat Reduction Agency (DTRA), C3.ai Digital Transformation Institute, Eric and Wendy Schmidt Center at the Broad Institute, Abdul Latif Jameel Clinic for Machine Learning in Health, DTRA Discovery of Medical Countermeasures Against New and Emerging (DOMANE) threats program, and DARPA Accelerated Molecular Discovery program.

\bibliography{iclr2022_conference}

\begin{thebibliography}{44}
\providecommand{\natexlab}[1]{#1}
\providecommand{\url}[1]{\texttt{#1}}
\expandafter\ifx\csname urlstyle\endcsname\relax
  \providecommand{\doi}[1]{doi: #1}\else
  \providecommand{\doi}{doi: \begingroup \urlstyle{rm}\Url}\fi

\bibitem[Abbas et~al.(2014)Abbas, Lichtman, and Pillai]{abbas2014cellular}
Abul~K Abbas, Andrew~H Lichtman, and Shiv Pillai.
\newblock \emph{Cellular and molecular immunology E-book}.
\newblock Elsevier Health Sciences, 2014.

\bibitem[Adolf-Bryfogle et~al.(2018)Adolf-Bryfogle, Kalyuzhniy, Kubitz,
  Weitzner, Hu, Adachi, Schief, and
  Dunbrack~Jr]{adolf2018rosettaantibodydesign}
Jared Adolf-Bryfogle, Oleks Kalyuzhniy, Michael Kubitz, Brian~D Weitzner,
  Xiaozhen Hu, Yumiko Adachi, William~R Schief, and Roland~L Dunbrack~Jr.
\newblock Rosettaantibodydesign (rabd): A general framework for computational
  antibody design.
\newblock \emph{PLoS computational biology}, 14\penalty0 (4):\penalty0
  e1006112, 2018.

\bibitem[Akbar et~al.(2021)Akbar, Robert, Weber, Widrich, Frank, Pavlovi{\'c},
  Scheffer, Chernigovskaya, Snapkov, Slabodkin, et~al.]{akbar2021silico}
Rahmad Akbar, Philippe~A Robert, C{\'e}dric~R Weber, Michael Widrich, Robert
  Frank, Milena Pavlovi{\'c}, Lonneke Scheffer, Maria Chernigovskaya, Igor
  Snapkov, Andrei Slabodkin, et~al.
\newblock In silico proof of principle of machine learning-based antibody
  design at unconstrained scale.
\newblock \emph{BioRxiv}, 2021.

\bibitem[Al~Qaraghuli et~al.(2020)Al~Qaraghuli, Kubiak-Ossowska, Ferro, and
  Mulheran]{al2020antibody}
Mohammed~M Al~Qaraghuli, Karina Kubiak-Ossowska, Valerie~A Ferro, and Paul~A
  Mulheran.
\newblock Antibody-protein binding and conformational changes: identifying
  allosteric signalling pathways to engineer a better effector response.
\newblock \emph{Scientific reports}, 10\penalty0 (1):\penalty0 1--10, 2020.

\bibitem[Alley et~al.(2019)Alley, Khimulya, Biswas, AlQuraishi, and
  Church]{alley2019unified}
Ethan~C Alley, Grigory Khimulya, Surojit Biswas, Mohammed AlQuraishi, and
  George~M Church.
\newblock Unified rational protein engineering with sequence-based deep
  representation learning.
\newblock \emph{Nature methods}, 16\penalty0 (12):\penalty0 1315--1322, 2019.

\bibitem[Baek et~al.(2021)Baek, DiMaio, Anishchenko, Dauparas, Ovchinnikov,
  Lee, Wang, Cong, Kinch, Schaeffer, et~al.]{baek2021accurate}
Minkyung Baek, Frank DiMaio, Ivan Anishchenko, Justas Dauparas, Sergey
  Ovchinnikov, Gyu~Rie Lee, Jue Wang, Qian Cong, Lisa~N Kinch, R~Dustin
  Schaeffer, et~al.
\newblock Accurate prediction of protein structures and interactions using a
  three-track neural network.
\newblock \emph{Science}, 373\penalty0 (6557):\penalty0 871--876, 2021.

\bibitem[Cao et~al.(2021)Cao, Das, Chenthamarakshan, Chen, Melnyk, and
  Shen]{cao2021fold2seq}
Yue Cao, Payel Das, Vijil Chenthamarakshan, Pin-Yu Chen, Igor Melnyk, and Yang
  Shen.
\newblock Fold2seq: A joint sequence (1d)-fold (3d) embedding-based generative
  model for protein design.
\newblock In \emph{International Conference on Machine Learning}, pp.\
  1261--1271. PMLR, 2021.

\bibitem[Cho et~al.(2014)Cho, Van~Merri{\"e}nboer, Gulcehre, Bahdanau,
  Bougares, Schwenk, and Bengio]{cho2014learning}
Kyunghyun Cho, Bart Van~Merri{\"e}nboer, Caglar Gulcehre, Dzmitry Bahdanau,
  Fethi Bougares, Holger Schwenk, and Yoshua Bengio.
\newblock Learning phrase representations using rnn encoder-decoder for
  statistical machine translation.
\newblock \emph{arXiv preprint arXiv:1406.1078}, 2014.

\bibitem[Dunbar et~al.(2014)Dunbar, Krawczyk, Leem, Baker, Fuchs, Georges, Shi,
  and Deane]{dunbar2014sabdab}
James Dunbar, Konrad Krawczyk, Jinwoo Leem, Terry Baker, Angelika Fuchs, Guy
  Georges, Jiye Shi, and Charlotte~M Deane.
\newblock Sabdab: the structural antibody database.
\newblock \emph{Nucleic acids research}, 42\penalty0 (D1):\penalty0
  D1140--D1146, 2014.

\bibitem[Fischman \& Ofran(2018)Fischman and Ofran]{fischman2018computational}
Sharon Fischman and Yanay Ofran.
\newblock Computational design of antibodies.
\newblock \emph{Current opinion in structural biology}, 51:\penalty0 156--162,
  2018.

\bibitem[Gebauer et~al.(2019)Gebauer, Gastegger, and
  Sch{\"u}tt]{gebauer2019symmetry}
Niklas~WA Gebauer, Michael Gastegger, and Kristof~T Sch{\"u}tt.
\newblock Symmetry-adapted generation of 3d point sets for the targeted
  discovery of molecules.
\newblock In \emph{Proceedings of the 33rd International Conference on Neural
  Information Processing Systems}, pp.\  7566--7578, 2019.

\bibitem[Grover et~al.(2019)Grover, Zweig, and Ermon]{grover2019graphite}
Aditya Grover, Aaron Zweig, and Stefano Ermon.
\newblock Graphite: Iterative generative modeling of graphs.
\newblock In \emph{International conference on machine learning}, pp.\
  2434--2444. PMLR, 2019.

\bibitem[Henikoff \& Henikoff(1992)Henikoff and Henikoff]{henikoff1992amino}
Steven Henikoff and Jorja~G Henikoff.
\newblock Amino acid substitution matrices from protein blocks.
\newblock \emph{Proceedings of the National Academy of Sciences}, 89\penalty0
  (22):\penalty0 10915--10919, 1992.

\bibitem[Ingraham et~al.(2018)Ingraham, Riesselman, Sander, and
  Marks]{ingraham2018learning}
John Ingraham, Adam Riesselman, Chris Sander, and Debora Marks.
\newblock Learning protein structure with a differentiable simulator.
\newblock In \emph{International Conference on Learning Representations}, 2018.

\bibitem[Ingraham et~al.(2019)Ingraham, Garg, Barzilay, and
  Jaakkola]{ingraham2019generative}
John Ingraham, Vikas~K Garg, Regina Barzilay, and Tommi Jaakkola.
\newblock Generative models for graph-based protein design.
\newblock \emph{Neural Information Processing Systems}, 2019.

\bibitem[Jin et~al.(2020{\natexlab{a}})Jin, Barzilay, and
  Jaakkola]{jin2020hierarchical}
Wengong Jin, Regina Barzilay, and Tommi Jaakkola.
\newblock Hierarchical generation of molecular graphs using structural motifs.
\newblock In \emph{Proceedings of the 37th International Conference on Machine
  Learning}, volume 119, pp.\  4839--4848. PMLR, 2020{\natexlab{a}}.

\bibitem[Jin et~al.(2020{\natexlab{b}})Jin, Barzilay, and
  Jaakkola]{jin2020multi}
Wengong Jin, Regina Barzilay, and Tommi Jaakkola.
\newblock Multi-objective molecule generation using interpretable
  substructures.
\newblock In \emph{Proceedings of the 37th International Conference on Machine
  Learning}, volume 119, pp.\  4849--4859. PMLR, 2020{\natexlab{b}}.

\bibitem[Jumper et~al.(2021)Jumper, Evans, Pritzel, Green, Figurnov,
  Ronneberger, Tunyasuvunakool, Bates, {\v{Z}}{\'\i}dek, Potapenko,
  et~al.]{jumper2021highly}
John Jumper, Richard Evans, Alexander Pritzel, Tim Green, Michael Figurnov,
  Olaf Ronneberger, Kathryn Tunyasuvunakool, Russ Bates, Augustin
  {\v{Z}}{\'\i}dek, Anna Potapenko, et~al.
\newblock Highly accurate protein structure prediction with alphafold.
\newblock \emph{Nature}, 596\penalty0 (7873):\penalty0 583--589, 2021.

\bibitem[Kabsch(1976)]{kabsch1976solution}
Wolfgang Kabsch.
\newblock A solution for the best rotation to relate two sets of vectors.
\newblock \emph{Acta Crystallographica Section A: Crystal Physics, Diffraction,
  Theoretical and General Crystallography}, 32\penalty0 (5):\penalty0 922--923,
  1976.

\bibitem[Karimi et~al.(2020)Karimi, Zhu, Cao, and Shen]{karimi2020novo}
Mostafa Karimi, Shaowen Zhu, Yue Cao, and Yang Shen.
\newblock De novo protein design for novel folds using guided conditional
  wasserstein generative adversarial networks.
\newblock \emph{Journal of Chemical Information and Modeling}, 60\penalty0
  (12):\penalty0 5667--5681, 2020.

\bibitem[Lapidoth et~al.(2015)Lapidoth, Baran, Pszolla, Norn, Alon, Tyka, and
  Fleishman]{lapidoth2015abdesign}
Gideon~D Lapidoth, Dror Baran, Gabriele~M Pszolla, Christoffer Norn, Assaf
  Alon, Michael~D Tyka, and Sarel~J Fleishman.
\newblock Abdesign: A n algorithm for combinatorial backbone design guided by
  natural conformations and sequences.
\newblock \emph{Proteins: Structure, Function, and Bioinformatics}, 83\penalty0
  (8):\penalty0 1385--1406, 2015.

\bibitem[Leaver-Fay et~al.(2011)Leaver-Fay, Tyka, Lewis, Lange, Thompson,
  Jacak, Kaufman, Renfrew, Smith, Sheffler, et~al.]{leaver2011rosetta3}
Andrew Leaver-Fay, Michael Tyka, Steven~M Lewis, Oliver~F Lange, James
  Thompson, Ron Jacak, Kristian~W Kaufman, P~Douglas Renfrew, Colin~A Smith,
  Will Sheffler, et~al.
\newblock Rosetta3: an object-oriented software suite for the simulation and
  design of macromolecules.
\newblock \emph{Methods in enzymology}, 487:\penalty0 545--574, 2011.

\bibitem[Lefranc et~al.(2003)Lefranc, Pommi{\'e}, Ruiz, Giudicelli, Foulquier,
  Truong, Thouvenin-Contet, and Lefranc]{lefranc2003imgt}
Marie-Paule Lefranc, Christelle Pommi{\'e}, Manuel Ruiz, V{\'e}ronique
  Giudicelli, Elodie Foulquier, Lisa Truong, Val{\'e}rie Thouvenin-Contet, and
  G{\'e}rard Lefranc.
\newblock Imgt unique numbering for immunoglobulin and t cell receptor variable
  domains and ig superfamily v-like domains.
\newblock \emph{Developmental \& Comparative Immunology}, 27\penalty0
  (1):\penalty0 55--77, 2003.

\bibitem[Lei(2021)]{lei2021attention}
Tao Lei.
\newblock When attention meets fast recurrence: Training language models with
  reduced compute.
\newblock \emph{arXiv preprint arXiv:2102.12459}, 2021.

\bibitem[Li et~al.(2014)Li, Pantazes, and Maranas]{li2014optmaven}
Tong Li, Robert~J Pantazes, and Costas~D Maranas.
\newblock Optmaven--a new framework for the de novo design of antibody variable
  region models targeting specific antigen epitopes.
\newblock \emph{PloS one}, 9\penalty0 (8):\penalty0 e105954, 2014.

\bibitem[Li et~al.(2018)Li, Vinyals, Dyer, Pascanu, and
  Battaglia]{li2018learning}
Yujia Li, Oriol Vinyals, Chris Dyer, Razvan Pascanu, and Peter Battaglia.
\newblock Learning deep generative models of graphs.
\newblock \emph{arXiv preprint arXiv:1803.03324}, 2018.

\bibitem[Liao et~al.(2019)Liao, Li, Song, Wang, Hamilton, Duvenaud, Urtasun,
  and Zemel]{liao2019efficient}
Renjie Liao, Yujia Li, Yang Song, Shenlong Wang, Will Hamilton, David~K
  Duvenaud, Raquel Urtasun, and Richard Zemel.
\newblock Efficient graph generation with graph recurrent attention networks.
\newblock \emph{Advances in Neural Information Processing Systems},
  32:\penalty0 4255--4265, 2019.

\bibitem[Liu et~al.(2020)Liu, Zeng, Mueller, Carter, Wang, Schilz, Horny,
  Birnbaum, Ewert, and Gifford]{liu2020antibody}
Ge~Liu, Haoyang Zeng, Jonas Mueller, Brandon Carter, Ziheng Wang, Jonas Schilz,
  Geraldine Horny, Michael~E Birnbaum, Stefan Ewert, and David~K Gifford.
\newblock Antibody complementarity determining region design using
  high-capacity machine learning.
\newblock \emph{Bioinformatics}, 36\penalty0 (7):\penalty0 2126--2133, 2020.

\bibitem[Liu et~al.(2018)Liu, Allamanis, Brockschmidt, and
  Gaunt]{liu2018constrained}
Qi~Liu, Miltiadis Allamanis, Marc Brockschmidt, and Alexander~L Gaunt.
\newblock Constrained graph variational autoencoders for molecule design.
\newblock \emph{Neural Information Processing Systems}, 2018.

\bibitem[Luo \& Hu(2021)Luo and Hu]{luo2021diffusion}
Shitong Luo and Wei Hu.
\newblock Diffusion probabilistic models for 3d point cloud generation.
\newblock In \emph{Proceedings of the IEEE/CVF Conference on Computer Vision
  and Pattern Recognition}, pp.\  2837--2845, 2021.

\bibitem[O'Connell et~al.(2018)O'Connell, Li, Hanson, Heffernan, Lyons,
  Paliwal, Dehzangi, Yang, and Zhou]{o2018spin2}
James O'Connell, Zhixiu Li, Jack Hanson, Rhys Heffernan, James Lyons, Kuldip
  Paliwal, Abdollah Dehzangi, Yuedong Yang, and Yaoqi Zhou.
\newblock Spin2: Predicting sequence profiles from protein structures using
  deep neural networks.
\newblock \emph{Proteins: Structure, Function, and Bioinformatics}, 86\penalty0
  (6):\penalty0 629--633, 2018.

\bibitem[Pantazes \& Maranas(2010)Pantazes and Maranas]{pantazes2010optcdr}
RJ~Pantazes and Costas~D Maranas.
\newblock Optcdr: a general computational method for the design of antibody
  complementarity determining regions for targeted epitope binding.
\newblock \emph{Protein Engineering, Design \& Selection}, 23\penalty0
  (11):\penalty0 849--858, 2010.

\bibitem[Pinto et~al.(2020)Pinto, Park, Beltramello, Walls, Tortorici, Bianchi,
  Jaconi, Culap, Zatta, De~Marco, et~al.]{pinto2020cross}
Dora Pinto, Young-Jun Park, Martina Beltramello, Alexandra~C Walls, M~Alejandra
  Tortorici, Siro Bianchi, Stefano Jaconi, Katja Culap, Fabrizia Zatta, Anna
  De~Marco, et~al.
\newblock Cross-neutralization of sars-cov-2 by a human monoclonal sars-cov
  antibody.
\newblock \emph{Nature}, 583\penalty0 (7815):\penalty0 290--295, 2020.

\bibitem[Raybould et~al.(2019)Raybould, Marks, Krawczyk, Taddese, Nowak, Lewis,
  Bujotzek, Shi, and Deane]{raybould2019five}
Matthew~IJ Raybould, Claire Marks, Konrad Krawczyk, Bruck Taddese, Jaroslaw
  Nowak, Alan~P Lewis, Alexander Bujotzek, Jiye Shi, and Charlotte~M Deane.
\newblock Five computational developability guidelines for therapeutic antibody
  profiling.
\newblock \emph{Proceedings of the National Academy of Sciences}, 116\penalty0
  (10):\penalty0 4025--4030, 2019.

\bibitem[Raybould et~al.(2021)Raybould, Kovaltsuk, Marks, and
  Deane]{raybould2021cov}
Matthew~IJ Raybould, Aleksandr Kovaltsuk, Claire Marks, and Charlotte~M Deane.
\newblock Cov-abdab: the coronavirus antibody database.
\newblock \emph{Bioinformatics}, 37\penalty0 (5):\penalty0 734--735, 2021.

\bibitem[Saka et~al.(2021)Saka, Kakuzaki, Metsugi, Kashiwagi, Yoshida, Wada,
  Tsunoda, and Teramoto]{saka2021antibody}
Koichiro Saka, Taro Kakuzaki, Shoichi Metsugi, Daiki Kashiwagi, Kenji Yoshida,
  Manabu Wada, Hiroyuki Tsunoda, and Reiji Teramoto.
\newblock Antibody design using lstm based deep generative model from phage
  display library for affinity maturation.
\newblock \emph{Scientific reports}, 11\penalty0 (1):\penalty0 1--13, 2021.

\bibitem[Shi et~al.(2021)Shi, Luo, Xu, and Tang]{shi2021learning}
Chence Shi, Shitong Luo, Minkai Xu, and Jian Tang.
\newblock Learning gradient fields for molecular conformation generation.
\newblock \emph{International Conference on Machine Learning}, 2021.

\bibitem[Shin et~al.(2021)Shin, Riesselman, Kollasch, McMahon, Simon, Sander,
  Manglik, Kruse, and Marks]{shin2021protein}
Jung-Eun Shin, Adam~J Riesselman, Aaron~W Kollasch, Conor McMahon, Elana Simon,
  Chris Sander, Aashish Manglik, Andrew~C Kruse, and Debora~S Marks.
\newblock Protein design and variant prediction using autoregressive generative
  models.
\newblock \emph{Nature communications}, 12\penalty0 (1):\penalty0 1--11, 2021.

\bibitem[Steinegger \& S{\"o}ding(2017)Steinegger and
  S{\"o}ding]{steinegger2017mmseqs2}
Martin Steinegger and Johannes S{\"o}ding.
\newblock Mmseqs2 enables sensitive protein sequence searching for the analysis
  of massive data sets.
\newblock \emph{Nature biotechnology}, 35\penalty0 (11):\penalty0 1026--1028,
  2017.

\bibitem[Strokach et~al.(2020)Strokach, Becerra, Corbi-Verge, Perez-Riba, and
  Kim]{strokach2020fast}
Alexey Strokach, David Becerra, Carles Corbi-Verge, Albert Perez-Riba, and
  Philip~M Kim.
\newblock Fast and flexible design of novel proteins using graph neural
  networks.
\newblock \emph{BioRxiv}, pp.\  868935, 2020.

\bibitem[Tischer et~al.(2020)Tischer, Lisanza, Wang, Dong, Anishchenko, Milles,
  Ovchinnikov, and Baker]{tischer2020design}
Doug Tischer, Sidney Lisanza, Jue Wang, Runze Dong, Ivan Anishchenko, Lukas~F
  Milles, Sergey Ovchinnikov, and David Baker.
\newblock Design of proteins presenting discontinuous functional sites using
  deep learning.
\newblock \emph{bioRxiv}, 2020.

\bibitem[Yang et~al.(2020{\natexlab{a}})Yang, Anishchenko, Park, Peng,
  Ovchinnikov, and Baker]{yang2020improved}
Jianyi Yang, Ivan Anishchenko, Hahnbeom Park, Zhenling Peng, Sergey
  Ovchinnikov, and David Baker.
\newblock Improved protein structure prediction using predicted interresidue
  orientations.
\newblock \emph{Proceedings of the National Academy of Sciences}, 117\penalty0
  (3):\penalty0 1496--1503, 2020{\natexlab{a}}.

\bibitem[Yang et~al.(2020{\natexlab{b}})Yang, Jin, Swanson, Barzilay, and
  Jaakkola]{yang2020improving}
Kevin Yang, Wengong Jin, Kyle Swanson, Regina Barzilay, and Tommi Jaakkola.
\newblock Improving molecular design by stochastic iterative target
  augmentation.
\newblock In \emph{International Conference on Machine Learning}, pp.\
  10716--10726. PMLR, 2020{\natexlab{b}}.

\bibitem[You et~al.(2018)You, Ying, Ren, Hamilton, and
  Leskovec]{you2018graphrnn}
Jiaxuan You, Rex Ying, Xiang Ren, William~L Hamilton, and Jure Leskovec.
\newblock Graphrnn: A deep generative model for graphs.
\newblock \emph{International Conference on Machine Learning}, 2018.

\end{thebibliography}
\bibliographystyle{iclr2022_conference}

\appendix
\newpage

\iffalse
\section{Antibody: a schematic illustration}

Figure~\ref{fig:antibody} gives a schematic structure of an antibody. It contains two identical heavy chains (blue, yellow) and two identical light chains (green, pink). A heavy chain contains one variable domain (VH) and three constant domains, while a light chain contains one variable domain (VL) and one constant domain. A VH sequence typically contains 100 and 130 residues, which are further divided into three CDRs and a framework region.

Most prior work~\citep{pantazes2010optcdr,adolf2018rosettaantibodydesign,liu2020antibody} focused on the design of CDR sequences. To verify that CDRs are the most important and variable part of an antibody, we trained a single-layer LSTM to generate the entire VH sequence. Indeed, the test perplexity of a full VH sequence is around 1.8, which is much lower than any of the CDR sequence perplexity (6.3$\sim$8.6).
\begin{figure}[h]
    \centering
    \vspace{-10pt}
    \includegraphics[width=0.9\textwidth]{figures/antibody.pdf}
    \caption{Schematic structure of an antibody (figure modified from Wikipedia).}
    \label{fig:antibody}
    \vspace{-10pt}
\end{figure}
\fi

\section{Model architecture details}

\subsection{RefineGNN}

\textbf{Node features.} Each node feature $\vv_i$ encodes three dihedral angles as follows.
\begin{equation}
    \vv_i = (\cos\phi_i, \cos\psi_i, \cos\omega_i, \sin\phi_i, \sin\psi_i, \sin\omega_i) \label{eq:node-feature}
\end{equation}
\textbf{Edge features.} The orientation matrix $\mO_i = [\vb_i, \vn_i, \vb_i \times \vn_i]$ defines a local coordinate system for each residue $i$ \citep{ingraham2019generative}, which is calculated as
\begin{equation}
    \vu_i = \frac{\vx_i - \vx_{i-1}}{\lVert \vx_i - \vx_{i-1} \rVert}, \quad
    \vb_i = \frac{\vu_i - \vu_{i+1}}{\lVert \vu_i - \vu_{i+1} \rVert}, \quad
    \vn_i = \frac{\vu_i \times \vu_{i+1}}{\lVert \vu_i \times \vu_{i+1} \rVert}
\end{equation}
\textbf{Attention mechanism.} The attention layer used in Eq.(\ref{eq:coarse}) is a standard bilinear attention:
\begin{equation}
    \mathrm{attention}(\vc_{1:n}, \vh_t) = \sum_i \alpha_{i,t} \vc_i, \qquad \alpha_{i,t} = \frac{\exp(\vc_i^\top \mW \vh_t)}{\sum_j \exp(\vc_j^\top \mW \vh_t)}
\end{equation}

\subsection{AR-GNN}
AR-GNN generates an antibody graph autoregressively. In each generation step $t$, AR-GNN learns to encode the current subgraph $\gG_{1:t}$ induced from residues $\set{\vs_1,\cdots,\vs_t}$ into a list of vectors
\begin{equation}
    \set{\vh_1,\cdots,\vh_t} = \MPN_{\theta}(\gG_{1:t}).
\end{equation}
For fair comparison, we use the same MPN architecture for both RefineGNN and AR-GNN. In terms of structure prediction, AR-GNN first predicts the node feature $\hat{\vv}_{t+1}$ of the next residue $t+1$, namely the dihedral angle between its three atoms $C_\alpha, C, N$.
\begin{equation}
    \hat{\vv}_{t+1} = \mW_v \vh_t
\end{equation}
In addition, AR-GNN predicts the pairwise distance between $\vs_{t+1}$ and previous residues $\vs_1,\cdots,\vs_t$.
\begin{equation}
    \hat{\vd}_{i,t+1} = \FFN(\mW_d \vh_i + \mU_d \vh_t + \mV_d E_{pos}(t + 1 - i)),
\end{equation}
where $\FFN$ is a feed-forward network with one hidden layer and $E_{pos}$ is the positional encoding of $t + 1 - i$, the \emph{gap} between residue $\vs_{t+1}$ and $\vs_i$ in the sequence. Lastly, AR-GNN predicts the amino acid type of residue $\vs_{t+1}$ by
\begin{equation}
    \hat{\vp}_{t+1} = \mathrm{softmax}(\mW_a \vg_{t+1}), \qquad \set{\vg_1,\cdots,\vg_{t+1}} = \MPN_{\theta'}(\gG_{1:t+1})
\end{equation}

Note that AR-GNN also uses two separate MPNs for structure and sequence prediction. However, unlike RefineGNN, AR-GNN is trained under teacher forcing --- we need to feed it the ground truth structure and sequence in each generation step. In particular, we find data augmentation to be crucial for AR-GNN performance. Data augmentation is essential because of the discrepancy between training and testing. The model is trained under teacher forcing, but it needs to decode a graph without teacher forcing at test time. We find mistakes made in previous steps have a great impact on subsequent predictions during decoding.

Specifically, for every antibody $\vs$, we create a \emph{corrupted} graph $\widetilde{\gG}$ by adding independent random Gaussian noise to every coordinate: $\tilde{\vx}_i = \vx_i + 3 \epsilon, \epsilon \sim \gN(0, I)$. In each generation step, we apply MPN over the corrupted graph instead.
\begin{equation}
    \set{\widetilde{\vh}_1,\cdots,\widetilde{\vh}_{t}} = \MPN_{\theta}(\widetilde{\gG}_{1:t}), \qquad \set{\widetilde{\vg}_1,\cdots,\widetilde{\vg}_{t+1}} = \MPN_{\theta'}(\widetilde{\gG}_{1:t+1})
\end{equation}
The node and edge labels are still defined by the ground truth structure. Specifically, let $\vv_t$ and $\vd_{i,j}$ be the ground truth dihedral angle and pairwise distance calculated from the original, uncorrupted graph $\gG$. AR-GNN loss function is defined as the following.
\begin{equation}
    \gL_{AR} = \sum_{i,j} \Vert \hat{\vd}_{i,j} - \vd_{i,j} \Vert^2 + \sum_t \Vert \hat{\vv}_t - \vv_t \Vert^2 + \gL_{ce}(\hat{\vp}_t, \vs_t)
\end{equation}
Similar to RefineGNN, AR-GNN also uses attention mechanism for conditional generation. Specifically, we concatenate the residue representations $\widetilde{\vh}_t, \widetilde{\vg}_t$ from MPN with context vectors learned from an attention layer.
\begin{equation}
    \widetilde{\vh}_t \leftarrow \widetilde{\vh}_t \oplus \mathrm{attention}(\vc_{1:n}, \widetilde{\vh}_t) \qquad
    \widetilde{\vg}_t \leftarrow \widetilde{\vg}_t \oplus \mathrm{attention}(\vc_{1:n}, \widetilde{\vg}_t)
\end{equation}

\section{Experimental details}

\textbf{Hyperparameters.} For AR-GNN and RefineGNN, we tried hidden dimension $d_h \in \set{128, 256}$ and number of message passing layers $L \in \set{1,2,3,4,5}$. We found $d_h=256,L=4$ worked the best for RefineGNN and $d_h=256,L=3$ worked the best for AR-GNN. For LSTM, we tried $d_h \in \set{128,256,512,1024}$. We found $d_h=256$ worked the best. All models are trained by an Adam optimizer with a dropout of 0.1 and a learning rate of 0.001.

\textbf{SAbDab data.} The dataset statistics of SAbDab is the following (after deduplication). For CDR-H1, the train/validation/test size is 4050, 359, and 326. For CDR-H2, the train/validation/test size is 3876, 483, and 376. For CDR-H3, the train/validation/test size is 3896, 403, and 437.

Since SAbDab includes both bound and unbound structures, we removed all antigens and used the bound antibody structure for training. Specifically, 65\% of our training data came from bound state structures. We included all data in our training set because the mismatch between bound and unbound structures is relatively small. In fact, \citet{al2020antibody} studied eight antibodies and found that the RMSD between bound and unbound structures over VH domains is less than 0.7 on average.

\textbf{RAbD configuration.} We provided details of the \emph{de novo} design setup of RosettaAntibodyDesign (RAbD) here. For each antibody in the test set, RAbD starts by randomly selecting a CDR from RAbD's internal database of known CDR structures. The chosen CDR-H3 sequence is required to have same length as the original sequence, but it cannot be exactly the same as the original CDR-H3 sequence.
After the initial CDR structure is chosen, RAbD grafts it onto the antibody and performs energy minimization to stabilize its structure. Next, RAbD runs 100 iterations of sequence design to modify the grafted CDR-H3 structure by randomly substituting amino acids. In each sequence design iteration, it performs energy minimization to adjust the structure according to the changed amino acid. Lastly, the model returns the generated CDR-H3 sequence with the lowest energy. 

\textbf{SARS-CoV-2 neutralization.}
Each generative model is pretrained on the SAbDab data to learn a prior distribution over CDR-H3 structures. Given a fixed predictor $f$, we use the ITA algorithm to finetune our pretrained models to generate neutralizing antibodies. Each model is trained for 3000 ITA steps with $M=100$. Generated CDR-H3 sequences from our model are visualized in Table~\ref{tab:examples}.

Our neutralization predictor $f$ is trained on the CoVAbDab database. For simplicity, we only consider two viruses, SARS-CoV-1 and SARS-CoV-2 since other coronavirus have very little training data. For the same reason, we only consider the spike protein receptor binding domain as our target epitope. The predictor is trained in a multi-task fashion to predict both SARS-CoV-1 and SARS-CoV-2 neutralization labels. The SRU encoder has a hidden dimension of 256. The model was trained with a dropout of 0.2, a learning rate of 0.0005, and batch size of 16.

The charge of a residue is defined as $C(\vs_i) = \mathbb{I}[\vs_i \in \{R, K\}] + 0.1 \cdot \mathbb{I}[\vs_i = H] - \mathbb{I}[\vs_i \in \{D, E\}]$ \citep{raybould2019five}. The net charge of a sequence $\vs_1\cdots\vs_n$ is defined as $\sum_i C(\vs_i)$.

\begin{table}
\centering
\caption{SARS-CoV-2 neutralization optimization results. Here we show examples of new CDR-H3 sequences generated by our model and their predicted neutralization improvement over original antibodies S1D7 and C694 in the CoVAbDab database.}
\label{tab:examples}
\begin{tabular}{lcc}
\hline
Antibody: & S1D7 & C694   \Tstrut\Bstrut \\
\hline
Old CDR-H3 & \texttt{TRGHSDY} & \texttt{ARDRGYDSSGPDAFDI}  \Tstrut\Bstrut\\
New CDR-H3 & \texttt{ARWWMDV} & \texttt{ARERIIIVSISAWMDV} \Tstrut\Bstrut\\
Improvement & 63\% $\rightarrow$ 73\% & 82\% $\rightarrow$ 91\% \Tstrut\Bstrut\\
\hline
\end{tabular}
\end{table}

\end{document}